\documentclass[runningheads]{llncs}
\usepackage{amsmath,amssymb}
\usepackage[T1]{fontenc}
\usepackage{graphicx}
\usepackage{hyperref}
\usepackage{color}

\usepackage{fancyvrb}

\usepackage{proof}

\newcommand{\seq}{\Rightarrow}
\newcommand{\sitz}{{\mathsf{unternehmenssitz\_in}}}
\newcommand{\standort}{{\mathsf{betriebsstandort\_in}}}
\newcommand{\rechtsform}{{\mathsf{rechtsform\_in}}}
\newcommand{\oenace}{{\mathsf{oenace\_in}}}

\DefineVerbatimEnvironment{Highlighting}{Verbatim}{commandchars=\\\{\}}
\newenvironment{Shaded}{}{}

\newcommand{\CommentTok}[1]{\textcolor[rgb]{0.38,0.63,0.69}{\textit{#1}}}

\newcommand{\DecValTok}[1]{\textcolor[rgb]{0.25,0.63,0.44}{#1}}

\newcommand{\KeywordTok}[1]{\textcolor[rgb]{0.00,0.44,0.13}{\textbf{#1}}}
\newcommand{\NormalTok}[1]{#1}
\newcommand{\OperatorTok}[1]{\textcolor[rgb]{0.40,0.40,0.40}{#1}}

\newcommand{\StringTok}[1]{\textcolor[rgb]{0.25,0.44,0.63}{#1}}

\begin{document}
\hyphenation{Unternehmens-register}
\hyphenation{Un-ter-neh-mens-ser-vice-por-tal}
\hyphenation{Unternehmens-sitz}
\hyphenation{Transparenz-portal}

\title{\emph{Grants4Companies}: Applying declarative methods for
  recommending and reasoning about business grants in the Austrian public
  administration (System description)}
\titlerunning{Grants4Companies (System description)}
%

\author{Bj{\"o}rn Lellmann\inst{1}\orcidID{0000-0002-5335-1838} \and
Philipp Marek\inst{2} \and 
Markus Triska\inst{1}}

\authorrunning{B. Lellmann, P. Marek, M. Triska}
%

\institute{Bundesministerium f{\"u}r Finanzen, Vienna, Austria\\
  \email{\{bjoern.lellmann, markus.triska\}@bmf.gv.at}
  \and Bundesrechenzentrum GmbH, Vienna, Austria \\
  \email{philipp.marek@brz.gv.at}}

\maketitle              
\begin{abstract}
  We describe the methods and technologies underlying the application
  \emph{Grants4Companies}. The application uses a logic-based expert system to
  display a list of business grants suitable for the logged-in business.
  To evaluate suitability of the
  grants, formal representations of their conditions are evaluated
  against properties of the business, taken from the registers of the
  Austrian public administration. The logical language for the representations of the
  grant conditions is based on S-expressions. We further describe a
  Proof of Concept implementation of reasoning over the formalised
  grant conditions. The proof of concept is implemented in Common Lisp
  and interfaces with a reasoning engine implemented in Scryer Prolog.
  The application has recently gone live and is provided as part of the
  \emph{Business Service Portal} by the Austrian Federal Ministry of Finance.
\end{abstract}
\section{Introduction}\label{introduction}
\label{sec:introduction}

Business grants are an important tool for steering and supporting the
economy. In addition, they can be used to quickly react to and counter
crises. However, the search for suitable business grants can be a
challenge for companies and businesses in Austria. This is due to the
large number of available business grants from a multitude of different
providers. While there are dedicated search engines, companies and
businesses often are simply not aware of the existence of grants on a
specific topic, and hence cannot use these engines in a targeted search.

As an additional tool for providing targeted information about potentially
interesting business grants to businesses the application
\emph{Grants4Companies} was introduced. The application is part of the
Austrian \emph{Unternehmensserviceportal}
(USP)\footnote{The official Austrian portal for interaction between
  businesses and public
  administration. See \url{https://www.usp.gv.at/en/ueber-das-usp/index.html}.} and is
productive since November
2022 with around 50 visits per month on average. The application uses data about available grants from the
Austrian \emph{Transparenzportal}\footnote{The official Austrian
  data base containing (amongst other information) data about the
  available grants. See \url{https://transparenzportal.gv.at}.} to
formalise formal grant conditions, e.g., on the type of business or the
location of the head office.
Data sources within the public administration are queried and used to evaluate these formalised criteria,
to display a list of grants ordered according to the feasibility of applying -
ie. whether the business fullfils the criteria or doesn't fullfil the
criteria; a third category contains grants for which the available information is not sufficient to decide.

While the application in the USP is written in Java\footnote{Due to interoperability concerns with existing libraries.}, we have also
implemented a Proof of Concept (PoC) for testing out new features, which we
describe in more detail in this article. In particular, this PoC
contains a reasoning engine for reasoning about the formalised grant
conditions themselves. The main features of the PoC are implemented in
Common Lisp while the reasoning engine is implemented in Scryer
Prolog\footnote{See \url{https://www.scryer.pl}}, following the Lean
Methodology~\cite{Beckert:1996} for implementing proof search in logical calculi
using Prolog's backtracking mechanism. The PoC is of interest for two
reasons: First, it combines implementations in Common Lisp and Scryer
Prolog to leverage the strengths of each programming language. Second,
it provides an example and showcase for the use of declarative
programming languages in public administration. To the best of our
knowledge, such examples are currently rather rare.

The source code for the reasoning engine complete with examples of
business grants with their conditions is available 
under \url{https://github.com/blellmann/g4c-reasoner}. 
While there is no openly accessible web interface, the reasoning
engine can be loaded into the \emph{Scryer Playground}\footnote{See \url{https://play.scryer.pl}}, the freely
accessible web interface for Scryer Prolog, and used for running evaluations.

In the remainder of the article we first give a brief overview of the
development history (Sec~\ref{sec:development-history}), followed by a description of the 
productive implementation of \emph{Grants4Companies}
(Sec.~\ref{sec:G4C-overview}) and the technical details underlying the
representation of the grants as well as their evaluation
(Sec.~\ref{sec:repr-and-eval}). We then provide details about the PoC
implementation (Sec.~\ref{sec:poc-ext}) including the implementation of the
reasoning engine and the interface between the Common Lisp
implementation and the Prolog reasoner, before concluding with an
outlook (Sec.~\ref{conclusion}). We do not include any benchmark results or
comparisons regarding efficiency of the reasoning engine here, since
the focus of the implementation is on correctness instead of maximal
efficiency, and it is part of a PoC implementation. Since the examples
of grants are taken from the official productive data set, we chose to
keep the original formulation of the examples and several concepts of
the representation language in German, providing additional
explanations in English. The technical terms from Austrian legislation
can of course be adapted to other languages.

\section{Development history}
\label{sec:development-history}

To assess the basic feasibility of the approach, we started with a
pilot~project, using Common~Lisp for rapid prototyping. Grants were
expressed as Lisp~forms, a natural representation when working with
Lisp. The pilot was successful, and also served as an illustration and
internal tool for communicating the approach we planned. Already in
this phase of the project, particular care was taken to explain in
the~UX that company data would only be processed with explicit
user~consent, and no data would be stored permanently by the planned
service. In order to demonstrate the key concepts without any
legal~concerns, the pilot did not use any real company data, but only
a fixed set of imaginary test companies.

For the production version of our service, we replaced the Common~Lisp
engine with a Java-based implementation to align the engine with
architectural principles of surrounding IT~services, and we retained
the representation of grants as Lisp~forms. As a result, the
Lisp-based pilot can still be seamlessly used on the production~data of
the formalised grants to
quickly prototype and assess additional features, while the Java-based
Lisp parser and evaluation engine can also be used in other
IT-services that require a Java implementation for architectural or
other reasons. Only the production version of the service has access
to real company data, and explicit consent of the company is required.

An additional component is the Prolog-based reasoner described in
Section~\ref{symbolic-reasoning-over-grants}. This component can be
used independently of the production environment to reason about
grants, and is freely provided in a public repository. This component
can reason with the productive formalised grants. Since the
reasoning concerns only logical relations between the grants
themselves, no company data is used by the reasoner.

\section{Grants4Companies Overview}
\label{sec:G4C-overview}

While the main focus of this article is the presentation of the PoC
implementation of extended features for \emph{Grants4Companies}, for
context we briefly describe the productive
application. \emph{Grants4Companies} is an application in the Austrian
\emph{Unternehmensserviceportal} (USP)\footnote{See \url{https://www.usp.gv.at/en/index.html}}. The USP is Austria's main
digital portal for the interaction between public administration and
businesses with currently more than 600.000 registered businesses and
more than 120 integrated applications. It also acts as identity provider for the businesses.

\begin{figure}[t]
  \hrule
  \centering
  \includegraphics[scale=0.5]{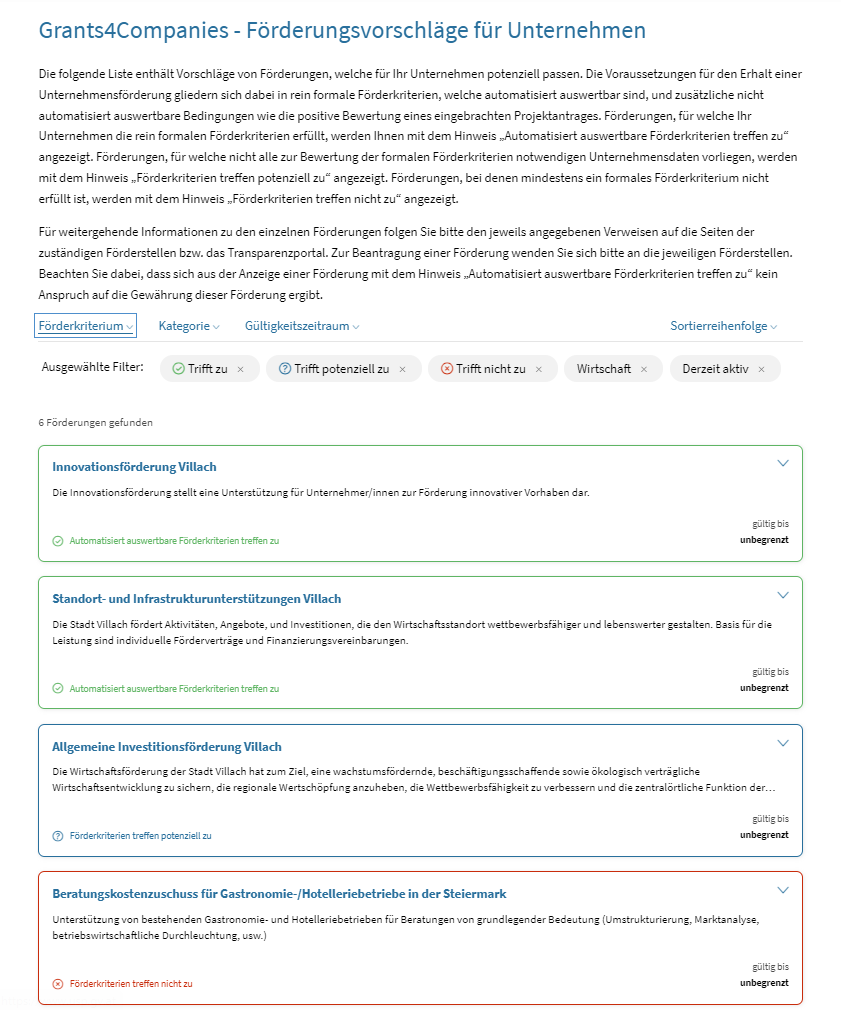}
  \hrule
  \caption{The productive version of \emph{Grants4Companies}. This
    figure shows the main page with a short description at the top and
  the list of grants. The grants are sorted with the applicable ones
  shown at the top of the list, the not applicable ones at the bottom,
and the ones requiring further data for a conclusive evaluation in the
middle. The list can be filtered, e.g., according to the topic of the
grant, in this case ``economy'' (``Wirtschaft'').}
  \label{fig:productive-g4c}
\end{figure}

After logging into the USP and starting the application
\emph{Grants4Companies}, businesses consent to the use of their data
from registers of public administration in line with the
GDPR~\cite{GDPR}. Following this consent, the application fetches
available data about the companies from registers of public
administration. Currently the data sources are the
\emph{Unternehmensregister} and the \emph{Firmenbuch}, the data used
concerns, e.g., information about the geographic location of the
business, its legal type, or the area of business following the
Austrian version of the NACE-classification\footnote{See \url{https://ec.europa.eu/eurostat/statistics-explained/index.php?title=Glossary:Statistical_classification_of_economic_activities_in_the_European_Community_(NACE)}}. The extension
to further registers is planned. Companies are then presented with a
list of grants, ordered according to whether the formal grant criteria
are satisfied by the company, not satisfied, or cannot be sufficiently
evaluated based on the available data. The latter option caters for
potential unavailability of necessary data from the registers, due to
lack of coverage or also maintenance downtime of the registers. The
results can be filtered and sorted according to the evaluation result,
categories of the grants, or application date. A screenshot of the productive version is shown in Fig.~\ref{fig:productive-g4c}.

The architecture of \emph{Grants4Companies} follows that of classical
knowledge based systems, with a clear separation between the knowledge
base, i.e., grant definitions including the formalised grant criteria,
and evaluation engine. The evaluation engine of the productive version
of \emph{Grants4Companies} is implemented in Java.
The knowledge base contains currently 45 grants which were formalised
manually. The details of the formal language used for representing the
grants will be considered in Sec.~\ref{selecting-a-language-for-specifying-grants}. 
The knowledge base is stored in a GIT
repository to keep track of historical data, and enable version control, reproducibility
and data sharing. This knowledge base is shared with the PoC
implementation.

\section{Representation and Evaluation of the Grant Conditions}
\label{sec:repr-and-eval}

The knowledge base containing the grants with their formalised grant
conditions is based on 
data about Austrian grants contained in the
Austrian
\emph{Transparenzportal}\footnote{See
  \url{https://transparenzportal.gv.at/tdb/tp/startpage} (in
  German).}, a portal provided by the Austrian Ministry of Finance,
where funding agencies are to enter 
grants and the granted
funding. For the PoC and the initial productive version of
\emph{Grants4Companies}, a number of grants were formalised
manually by us, the current knowledge base contain 45 grants. In the future
this might be extended following a rules as
code approach~\cite{Mowbray:2023CLSR}, e.g., using tools like \emph{POTATO}~\cite{Kovacs:2022Potato,Recski:2021ASAIL} for automatically suggesting formalised
grant conditions based on the natural language descriptions provided
by the funding agencies.

\subsection{Representation of the grants}\label{selecting-a-language-for-specifying-grants}

The grant conditions are formalised as quantifier-free logical
formulae. The language contains predicates for expressing properties
of the businesses related to location, legal form, classification of
business activity, etc. Examples of \emph{atomic formulae} with their
intended semantics are given in Fig.~\ref{fig:atomic-formulae}.
\begin{figure}[t]
\hrule\smallskip
\begin{center}
\begin{tabular}{l@{\quad}p{7cm}}
  Atomic Formula & Intended semantics\\
  \hline
  \texttt{Betriebsstandort-in(L)} & The business has a location in one
                                    of the areas/regions specified in
                                    the list \texttt{L}\\
  \texttt{Rechtsform-in(L)} & The legal form of the business is one of
                           those in the
                           list \texttt{L}\\
  \texttt{{\"O}NACE-in(L)} & The business activity classification
                             falls under one of the areas in the list \emph{L}
\end{tabular}
\end{center}
\hrule
\caption{Examples of atomic formulae and their intended semantics}
\label{fig:atomic-formulae}
\end{figure}
For ease of use by Austrian funding agencies,
these predicates are formulated in German and often take a list as
argument. \emph{Complex formulae} are built from the
atomic formulae as well as $\top,\bot$ as usual using the standard propositional
connectives $\neg, \lor, \land, \to$. At the current state there was
no need for quantifiers, these might be added in the future. Working
in a quantifier-free language has the benefit of a greatly reduced
complexity for the reasoning tasks, of course. For the sake of referring to
commonly used concepts, the language also contains \emph{defined
  concepts}. On the logical level, these are given as pairs
$(\mathfrak{d},D)$ consisting of the name $\mathfrak{d}$ of the
concept, which can be used like an atomic formula, and its definition
$D$, i.e., a formula not containing $\mathfrak{d}$. The
definition might contain other defined concepts, absence of cycles
is assumed to be ensured externally. E.g., the concept of a legal
person is introduced as the an abbreviation with name \texttt{G4c/Grants\_Gv.At:Ist-Juristische-Person} for the formula
\texttt{Rechtsform-in(L)}, where $L = [ \mathtt{Genossenschaft},
\mathtt{Verein}, \dots ]$ is a list of the legal forms
which count as legal persons in Austria.
Naming the definitions in the style of packages makes it
possible to differentiate between concepts with the same name
from different funding agencies, e.g., general funding conditions
specific to the funding agencies.

On a technical level, the language used for representing the logical
formulae is based on the Lisp-syntax of
\emph{S-expressions}~\cite[102]{belzer1978encyclopedia}. In
particular, the logical formulae formalising the grant conditions are
represented in prefix notation as lists, where the first element is
the logical connective and the following elements are its
arguments. E.g., a formula $\neg A \land (B \lor C)$ is represented as
the S-expression \texttt{(and (neg A) (or B C))}. Predicates are
represented by (Common Lisp) symbols. E.g., the predicate 
\texttt{Betriebsstandort-in} represents the fact that the business has
a location in one of a list of certain areas given by their
\emph{Gemeindekennzahl}, the Austrian identification number for
municipalities. To enable restriction also on a regional or county
level, also prefixes of these identification numbers are
covered. E.g., the atomic formula \texttt{(Betriebsstandort-in 2 617
  60101)} represents the assertion that the business has a site in
the county Carinthia, the region East Styria, or the municipality of
the city of Graz.

The full representation of a grant also contains in addition to the
formalised grant conditions also its name, metadata about application
dates and links to the full description on the Transparenzportal, as
well as the natural language description of the grant conditions. The
latter are included as Lisp comments interspersed with the formalised
conditions in the spirit of literate programming~\cite{Knuth:1984CJ}.
This allows to have human-readable explanations collected and used for
explaining the evaluation of a grant. An example of a grant in this
representation is given in Fig.~\ref{fig:ex-grant}. Defined concepts
$(\mathfrak{d},D)$ are represented as \texttt{(def-concept
  $\mathfrak{d}$ $D$)}.
An issue that came up right from the beginning is having one concept in
multiple different implementations. A clause specifying that the
company has to be a small or medium enterprise (SME, in German
``\emph{Der Antragsteller muss ein KMU 
sein}'') is used in many grants; sadly
there are three different definitions for this term, one from the
federal government in Austria, one from the EU, and one from the
FFG\footnote{The \emph{Österreichische Förderagentur für wirtschaftsnahe
  Forschung, Entwicklung und Innovation}, in English \emph{Austrian
  Research Promotion Agency}.}.
As mentioned, this ambiguity is solved via package names - there
are simply three functions, \texttt{GV.AT:IS-KMU}, \texttt{FFG:IS-KMU},
and \texttt{EU:IS-KMU}. This enables the use of different
interpretations of the same natural language term depending on the
source of the regulation.
An example of a defined concept is given in Fig.~\ref{fig:def-concept}.
\begin{figure}[t]
  \hrule\smallskip
  \texttt{
 (def-concept gv.at:natürliche-oder-juristische-Person\\
\phantom{.}\quad   (OR\\
\phantom{.}\quad\quad     (Rechtsform-in :Einzelunternehmen)\\
\phantom{.}\quad\quad     (gv.at:Ist-Juristische-Person)))}\smallskip
  \hrule
  \caption{The definition of the concept
    \texttt{gv.at:natürliche-oder-juristische-Person}. The formula
    captures the condition that the applicant is a natural person,
    i.e., the legal form of the company is that of a sole trader
    (\texttt{:Einzelunternehmen}), or a legal person (captured by the
    defined formula \texttt{gv.at:Ist-Juristische-Person}).}
  \label{fig:def-concept}
\end{figure}

\begin{figure} 
  \hrule\medskip
\begin{Shaded}
\begin{Highlighting}[]
\NormalTok{(define{-}grant (}\StringTok{"Umweltschutz{-} und Energieeffizienzförderung {-} Förderung}\\
\StringTok{ sonstiger Energieeffizienzmaßnahmen Villach"}
\NormalTok{   (:href }\StringTok{"https://transparenzportal.gv.at/tdb/tp/leistung/1052703.html"}\NormalTok{)}
\NormalTok{   (:transparenzportal{-}ref{-}nr }\DecValTok{1052703}\NormalTok{)}
\NormalTok{   (:Fördergebiet :Umwelt)}
\NormalTok{   (gültig{-}von }\StringTok{"2019{-}01{-}01"}\NormalTok{))}
  \StringTok{"Unter der Berücksichtigung der Verwendung erneuerbarer Energieträger }
\StringTok{   sowie der Umsetzung der Intention der Umweltschutz{-} und }
\StringTok{   Energieeffizienzrichtlinie im Bereich privater Haushalte fördert die }
\StringTok{   Stadt Villach folgende Energieeffizienzmaßnahmen."}
  \CommentTok{;; Voraussetzungen}
  \CommentTok{;;}
  \CommentTok{;; {-} Förderungswerber/innen können natürliche oder juristische Personen}
  \CommentTok{;;   sein. Bei juristischen Personen hat die firmenmäßige bzw. }
  \CommentTok{;;   statutenkonforme Unterfertigung des Antrages auf Gewährung einer }
  \CommentTok{;;   Förderung durch den Vertretungsbefugten zu erfolgen.}
\NormalTok{  (AND}
\NormalTok{    (GV.AT:natürliche{-}oder{-}juristische{-}Person)}
    \CommentTok{;; {-} Die Förderungswerber haben bei der Antragstellung zu erklären, }
    \CommentTok{;;   dass für die beantragten Förderungen keine weiteren Förderungen }
    \CommentTok{;;   von anderen Stellen beantragt wurden.}
    \CommentTok{;; {-} Ein Förderungsansuchen muss spätestens innerhalb von 8 Monaten }
    \CommentTok{;;   nach Umsetzung der Maßnahme/n bzw. Kaufdatum bei der Stadt}
    \CommentTok{;;   Villach einlangen}
    \CommentTok{;; {-} Die Förderung wird nur für die sach{-} und fachgerechten Umsetzung }
    \CommentTok{;;   der Maßnahme (Einbau) im Stadtgebiet von Villach gewährt.}
\NormalTok{    (OR}
\NormalTok{      (Unternehmenssitz{-}in }\DecValTok{20201}\NormalTok{)}
\NormalTok{      (Betriebsstandort{-}in }\DecValTok{20201}\NormalTok{))))}
\end{Highlighting}
\end{Shaded}
\hrule

\caption{Example grant, TP-Nr.1052703. The grant provides funding for
  increasing energy efficiency. It is applicable to natural and legal
  persons (\texttt{GV.AT:natürliche-oder-juristische-Person}) in the
  city of Villach (the \texttt{Unternehmenssitz} or a
  \texttt{Betriebsstandort} has to be in the municipal identification number
  20201). Some other conditions  cannot be checked
  automatically based on the data about the company available within
  public administration and hence are not formalised (e.g., that the request for
  funding has to be submitted at most 8 months after implementing the
  measures for increasing energy efficiency).}
\label{fig:ex-grant}
\end{figure}

\begin{figure} 
\hrule\medskip

\begin{Shaded}
\begin{Highlighting}[]
\NormalTok{förderung}\OperatorTok{(}\StringTok{"G4c/Grants\_Umweltschutz{-} Und Energieeffizienzförderung}\\
\StringTok{ {-} Förderung Sonstiger Energieeffizienzmaßnahmen Villach"}\OperatorTok{,}
\NormalTok{  förderkriterien}\OperatorTok{(}
    \OperatorTok{(}
\NormalTok{      df}\OperatorTok{(} \StringTok{"G4c/Grants\_Gv.At:Natürliche{-}Oder{-}Juristische{-}Person"} \OperatorTok{)}
\NormalTok{    and}
      \OperatorTok{(}
\NormalTok{        at}\OperatorTok{(}\NormalTok{ unternehmenssitz\_in}\OperatorTok{(} \OperatorTok{[} \DecValTok{20201} \OperatorTok{]} \OperatorTok{)} \OperatorTok{)}
\NormalTok{      or}
\NormalTok{        at}\OperatorTok{(}\NormalTok{ betriebsstandort\_in}\OperatorTok{(} \OperatorTok{[} \DecValTok{20201} \OperatorTok{]} \OperatorTok{)} \OperatorTok{)}
      \OperatorTok{)}
    \OperatorTok{)}
  \OperatorTok{)}
\OperatorTok{).}
\end{Highlighting}
\end{Shaded}
\hrule
\caption{Example grant in Prolog syntax, TPPNr\#1052703. For the
  original formulation of this particular grant, see Fig.~\ref{fig:ex-grant}}
\label{fig:ex-grant-prolog}
\end{figure}

\subsection{Evaluation of the grants}\label{evaluating}

Evaluating whether the formal conditions of a grant apply for a
specific business essentially corresponds to checking, whether the
business is a model of the logical formula representing these
conditions. Here the business is identified with its properties given
by the data about the business available. The atomic formulae are
chosen to directly correspond to data fields from specific registers
and hence their evaluation is rather straightforward. Complex formulae
are evaluated according to their main logical connectives. Names
$\mathfrak{d}$ for defined concepts $(\mathfrak{d},D)$ are unpacked
into their definition $D$
and then evaluated.

Of course
not
all the data required to evaluate whether a company satisfies the
formalised eligibility criteria of a grant is necessarily always available. While data like
\emph{location} of a company needs to be provided before it is
officially recognized, e.g., the (Ö)NACE
classification\footnote{\url{https://www.statistik.at/en/databases/classification-database}}
of the economic activities of Austrian businesses
is not complete. 
In particular, for a sizeable number of companies the
{\"O}NACE-classification has not yet been assigned. In addition, the
connection to a specific register might drop out temporarily due to
maintenance work.

To cover these eventualities, the evaluation is done in a three-valued
logic, which allows a third truth value of \emph{unknown} next to
\emph{true} and \emph{false}. The logical connectives then propagate
the truth value \emph{unknown} upwards, whenever no definite
evaluation to \emph{true} or \emph{false} is possible. 
To be precise, we use (so far
quantifier-free) \emph{strong Kleene-Logic} $K_3$,
considered, e.g., in~\cite{Kleene:1952fk}. The truth tables for the
logical connectives are given in
Fig.~\ref{fig:kleene-logic}. This ensures that grants
which have been evaluated for a company to \emph{true} or \emph{false} while
some of their atomic components are evaluated to \emph{unknown}
are evaluated with the same result when additional data becomes
available and some of the atomic components are no longer evaluated as \emph{unknown}.
Range-based reasoning for numeric
operations would also be possible, and is planned as future work. 
\begin{figure}[t]
\hrule\smallskip
\[
\begin{array}{c|ccc}
   & \neg \\
  \hline
  \bot & \top \\
  u & u \\
  \top & \bot 
\end{array}
\qquad
\begin{array}{c|ccc}
  \land & \bot & u & \top\\
  \hline
  \bot & \bot & \bot & \bot \\
  u & \bot & u & u \\
  \top & \bot & u & \top
\end{array}
\qquad
\begin{array}{c|ccc}
  \lor & \bot & u & \top\\
  \hline
  \bot & \bot & u & \top \\
  u & u & u & \top \\
  \top & \top & \top & \top
\end{array}
\qquad
\begin{array}{c|ccc}
  \to & \bot & u & \top\\
  \hline
  \bot & \top & \top & \top \\
  u & u & u & \top \\
  \top & \bot & u & \top 
\end{array}
\]
\hrule
\caption{The truth tables for 3-valued strong Kleene logic $K_3$. The
  truth values \emph{false}, \emph{unknown}, \emph{true} are
  represented by $\bot$, $u$ and $\top$, respectively.}
\label{fig:kleene-logic}
\end{figure}

As a further potential next step, the symbolic representation also allows for some easy optimizations --
for commutative connectives/operations (like \texttt{AND},
\texttt{OR}, possibly in the future also numerical
addition via \texttt{+}), we could reorder the arguments 
before evaluating. 
By moving the subformula with the highest probability for a negative
result to the front, a short-cutting evaluation could quickly discard
grant/company pairs, allowing for mass assessments: given a newly
proposed grant, how many companies in Austria will (be able to) apply?
This reordering is not implemented yet, though.

\section{PoC: Extensions and Interfaces}
\label{sec:poc-ext}

The PoC also contains an implementation of the evaluation of grant
conditions based on company data. However, for the purpose of this
article we concentrate on the functionality which goes beyond that of
the productive system. In particular, the representation of grand
conditions as logical formulae opens the possibility to not just
evaluate the conditions based on business data, but to also reason
\emph{about} the conditions themselves. Interesting questions here are
in particular consistency, useful for discovering mistakes in the
formalisation of grant conditions, and logical implication, useful for
finding unintended overlap between multiple grants in the same
area. To enable such reasoning, we implemented backwards proof search
in a Gentzen-style sequent calculus (see, e.g.,~\cite{Troelstra:2000fj} for the
proof-theoretic background). Following the
Lean-methodology~\cite{Beckert:1996} we make use of Prolog's
backtracking mechanism to perform the proof search.

We use Scryer Prolog due to its strong conformance to the Prolog
ISO~standard, which will ease future cooperations with other
organizations and public administrations. In addition, the system is
freely available and allows inspection of its entire source code,
which works towards our aim of providing full transparency and
explainability of all computed results.

\subsection{Symbolic Reasoning over
Grants}\label{symbolic-reasoning-over-grants}

The Proof-of-Concept has the ability to connect one or more \emph{Scryer
Prolog}\footnote{See \url{www.scryer.pl}} sessions to the web frontend, providing a convenient REPL
that is
pre-loaded with some known facts and the transpiled grant forms. We
included a prototypical implementation of logical reasoning over the
formalised eligibility criteria in the form of a \emph{sequent
  calculus}, specifically a \textsf{G3}-style calculus for classical
(propositional) logic (see, e.g.~\cite{Troelstra:2000fj}), extended to
cover basic facts about atomic statements and the defined concepts. We
use reasoning in classical logic and not the three-valued logic used
for evaluating the grants, because reasoning about the logical
properties of grant conditions is independent of the data available
for particular businesses. A calculus for the three-valued logic used
could be implemented, e.g.,
following~\cite{multlog:stongKleene}. However, this would be useful
mainly for reasoning about which grants are shown to the business with
which evaluation.

As usual, \emph{sequents} are of the form $A_1, \dots, A_n \seq B_1, \dots, B_m$
with $n,m \geq 0$ and are interpreted as the logical formula $A_1
\land \dots \land A_n \to B_1 \lor \dots \lor B_m$.
The standard logical
rules are given in Fig.~\ref{fig:sequent-calculus}. Basic knowledge
about implications between atomic statements is included in the form
of (rather simple) \emph{ground sequents}, and defined concepts
are included in the form of separate left- and right rules for each
defined concept. The ground sequents and
rules for unpacking the defined concepts are given in
Fig.~\ref{fig:ground-sequents}. Cut-free completeness of the calculus
follows from an extension of~\cite[Thm.4.6.1]{Troelstra:2000fj} to the
calculus with defined formulae, noting that the set of ground sequents
is closed under substitutions (because no variables occur),
contraction and basic cuts.
\begin{figure}[t]
\hrule\medskip
\[
  \infer[]{\Gamma, A \seq A, \Delta}{}
  \qquad
  \infer[\bot_L]{\Gamma, \bot \seq \Delta}{}
  \qquad
  \infer[\top_R]{\Gamma \seq \top, \Delta}{}
\]
\[
  \infer[\neg_L]{\Gamma, \neg A \seq \Delta
  }
  {\Gamma \seq A, \Delta
  }
  \qquad
  \infer[\neg_R]{\Gamma \seq \neg A, \Delta
  }
  {\Gamma, A \seq \Delta
  }
\]
\[
  \infer[\land_L]{\Gamma, A \land B \seq \Delta
  }
  {\Gamma, A, B \seq \Delta
  }
  \qquad
  \infer[\land_R]{\Gamma \seq A \land B, \Delta
  }
  {\Gamma \seq A, \Delta
    &
    \Gamma \seq B, \Delta
  }
\]
\[
  \infer[\lor_L]{\Gamma, A \lor B \seq \Delta
  }
  {\Gamma, A \seq \Delta
    &
    \Gamma, B \seq \Delta
  }
  \qquad
  \infer[\lor_R]{\Gamma \seq A \lor B, \Delta
  }
  {\Gamma \seq A, B \Delta
  }
\]
\[
  \infer[\to_L]{\Gamma, A \to B \seq D
  }
  {\Gamma, B \seq \Delta
    &
    \Gamma \seq A, \Delta
  }
  \qquad
  \infer[\to_R]{\Gamma \seq A \to B, \Delta
  }
  {\Gamma, A \seq B, \Delta
  }
\]
\hrule
\caption{The sequent rules of the propositional part of calculus \textsf{G3}}
\label{fig:sequent-calculus}
\end{figure}
\begin{figure}[t]
  \hrule
  \[
    \infer[\begin{array}{l}\forall x \in L_1 \exists y \in L_2: \\
             \quad y
             \mathrm{\;is\; prefix\; of\;}x \end{array}]{\Gamma, \sitz( L_1 ) \seq \sitz(L_2), \Delta
    }
    {
    }
  \]
  \[
    \infer[\begin{array}{l}\forall x \in L_1 \exists y \in L_2: \\
             \quad y
             \mathrm{\;is\; prefix\; of\;}x \end{array}]{\Gamma, \standort( L_1 ) \seq \standort(L_2), \Delta
    }
    {
    }
  \]
  \[
    \infer[\begin{array}{l}\forall x \in L_1 \exists y \in L_2: \\
             \quad y
             \mathrm{\;is\; prefix\; of\;}x \end{array}]{\Gamma, \oenace( L_1 ) \seq \oenace(L_2), \Delta
    }
    {
    }
  \]
  \[
    \infer[L_1 \subseteq L_2]{\Gamma, \rechtsform( L_1 ) \seq \rechtsform(L_2), \Delta
    }
    {
    }
  \]
  \[
    \infer[(\mathfrak{d}, D)_L 
    ]{\Gamma, \mathfrak{d} \seq \Delta
    }
    {\Gamma, D \seq \Delta
    }
    \qquad\qquad
    \infer[(\mathfrak{d}, D)_R 
    ]{\Gamma
      \seq \mathfrak{d}, \Delta
    }
    {\Gamma \seq D, \Delta
    }
  \]
  \hrule
  \caption{The ground sequents and definition rules used in the
    calculus. In the definition rules the pair $(\mathfrak{d},D)$ is a defined concept.}
  \label{fig:ground-sequents}
\end{figure}
In order to avoid unnecessary repetitions, in the implementation the rules
are given as facts about the term \texttt{rule(Name, Prem\_List / PF)}, where
\texttt{Name} is the name of the rule, \texttt{Prem\_List} is the list
of premisses and \texttt{PF} is the prinicipal formula of the rule,
i.e., a sequent with exactly one formula on the left or right hand
side. The provability predicate is given by \texttt{prov2//2}, which
is true if the first argument is a derivable sequent, and the second
argument a term describing a corresponding derivation. Examples are given in
Fig.~\ref{fig:code-rules}.
\begin{figure}[t]
  \hrule\smallskip
  \texttt{rule(andL, [[A,B]=>[]] / and(A,B)=>[]).}\\
  \texttt{rule(andR, [ [] => [A], [] => [B] ] / [] =>
    and(A,B)).}\medskip\\
  \texttt{prov2(Gamma => Delta, der(Rule\_name, Gamma => Delta, [Fml] => [], New\_prems\_ders)) :-\\
    \phantom{.}\quad select(Fml,Gamma,Omega),\\
    \phantom{.}\quad rule(Rule\_name,Prems / Fml => []),\\
    \phantom{.}\quad merge\_sequent\_list(Prems, Omega => Delta, New\_Prems),\\
    \phantom{.}\quad maplist(prov2,New\_Prems, New\_prems\_ders).}\medskip\\
  \texttt{
    prov2(Gamma => Delta, der(Rule\_name, Gamma => Delta, [] => [Fml], New\_prems\_ders)) :-\\
    \phantom{.}\quad select(Fml, Delta, Pi),\\
    \phantom{.}\quad rule(Rule\_name, Prems / [] => Fml),\\
    \phantom{.}\quad merge\_sequent\_list(Prems,Gamma => Pi, New\_prems),\\
    \phantom{.}\quad maplist(prov2, New\_prems, New\_prems\_ders).}\smallskip
  \hrule
  \caption{Examples of the Prolog code for rules of the sequent calculus.}
  \label{fig:code-rules}
\end{figure}
The auxiliary predicate
\texttt{merge\_sequent\_list//3} is true if the first argument
contains a list $\Gamma_1 \seq \Delta_1,\; ... , \Gamma_n \seq
\Delta_n$ of premisses, i.e., sequents, the second argument
contains a sequent $\Sigma \seq \Pi$, and the third argument contains
the list of premisses merged with this sequent, i.e., $\Gamma_1,\Sigma
\seq \Pi, \Delta_1,\; \dots, \Gamma_n, \Sigma \seq \Pi, \Delta_n$.
Since the rules of the calculus are invertible, we could introduce
prolog cuts \texttt{!} after the goals \texttt{rule(Rule\_name, ...)}
to increase efficiency -- this would not
influence completeness wrt.\ derivability of sequents. However, since this
\emph{would} limit the number of derivations found, and to preserve
monotonicity of the program, we refrain from doing so. 

The result of querying
for logical implication between the formalised eligibility conditions
of two grants is shown in Fig.~\ref{fig:reasoning}. The prolog
variable \texttt{Tree} is instantiated with a term for the derivation
of the result abbreviated here for the sake of better readability.

\begin{figure}
\hrule\medskip
\small
\begin{Shaded}
\begin{Highlighting}[]
\NormalTok{\% förderung(F1, förderkriterien(K1)), }
\NormalTok{  förderung(F2, förderkriterien(K2)),}
\NormalTok{  dif(K1, K2),}
\NormalTok{  prov2([K1] => [K2], Tree).}
\NormalTok{F1 }\OperatorTok{=} \StringTok{"Per{-}Bundesland/Steiermark\_Beratungskostenzuschuss{-}Für{-}Gastronomie{-}}
\StringTok{      /Hotelleriebetriebe{-}In{-}Der{-}Steiermark"}\NormalTok{, }
\NormalTok{K1 }\OperatorTok{=}\NormalTok{ ( df(}\StringTok{"Gv.At\_Natürliche{-}Oder{-}Juristische{-}Person"}\NormalTok{)}
       \KeywordTok{or} 
\NormalTok{       at(rechtsform\_in([}\StringTok{"Offene{-}Gesellschaft"}\NormalTok{,}\StringTok{"Kommanditgesellschaft"}\NormalTok{]))}
\NormalTok{     )}
     \KeywordTok{and}
\NormalTok{     ( at(önace\_in([}\StringTok{"55"}\NormalTok{]))}
       \KeywordTok{or} 
\NormalTok{       at(önace\_in([}\StringTok{"56"}\NormalTok{]))}
\NormalTok{     )}
     \KeywordTok{and}
\NormalTok{     ( at(unternehmenssitz\_in([}\StringTok{"Land{-}Stmk"}\NormalTok{]))}
       \KeywordTok{or} 
\NormalTok{       at(betriebsstandort\_in([}\StringTok{"Land{-}Stmk"}\NormalTok{]))}
\NormalTok{     ),}
\NormalTok{F2 }\OperatorTok{=} \StringTok{"Per{-}Bundesland/Steiermark\_Förderung{-}Zur{-}Wirtschaftsinitiative{-}}
\StringTok{      Nachhaltigkeit{-}Steiermark"}\NormalTok{, }
\NormalTok{K2 }\OperatorTok{=}\NormalTok{ at(unternehmenssitz\_in([}\StringTok{"Land{-}Stmk"}\NormalTok{]))}
     \KeywordTok{or} 
\NormalTok{     at(betriebsstandort\_in([}\StringTok{"Land{-}Stmk"}\NormalTok{])),}
\NormalTok{Tree }\OperatorTok{=}\NormalTok{ der(andL,...)}
\end{Highlighting}
\end{Shaded}
\hrule
\caption{Reasoning over grants. The query is shown at the top. The
  variables \texttt{F1} and \texttt{F2} are instantiated with the names of grants,
  such that the conditions \texttt{K1} of the first one imply the conditions
  \texttt{K2} of the second one. The variable \texttt{Tree} is instantiated
  with the  derivation witnessing provability of the sequent $\mathtt{K1} \seq
  \mathtt{K2}$, abbreviated here for the sake of space.}
\label{fig:reasoning}
\end{figure}

\begin{figure}[t]
  \hrule
  \centering
	\includegraphics[scale=0.4]{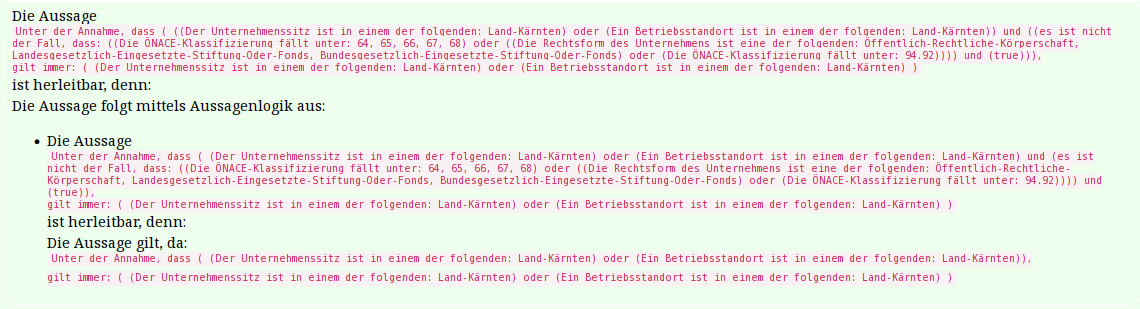}
  \hrule
  \caption{HTML output of a Prolog reasoning.}
	\label{fig:html-prolog}
\end{figure}

The terms representing derivations can also be converted into
human-readable form in a formalised natural language using Prolog's
Definite Clause Grammars. The result is a string containing html-code
which can be displayed in a browser, see Fig.~\ref{fig:html-prolog}.

\subsection{Interface between Lisp and Prolog}
\label{sec:lisp-prolog}

To enable the reasoning functionality from within the Lisp-part of the
PoC, the prolog prover is called and its output on \texttt{std\_out}
interpreted.  For this, the implementation of a basic interface was
necessary.

\subsubsection{Conversion from grant-code in Lisp-syntax to Prolog.}

Since the formalised grant conditions are given in the syntax of
S-Expressions, they need to be converted to Prolog terms.
While there is an existing project that transpiles S-expressions to ISO Prolog\footnote{https://github.com/cl-model-languages/cl-prolog2},
it didn't fit our usecase; this library only allows batch processing and not the desired interactive querying, the already-parsed internal grant structure isn't supported,
and a few special-cases demand a non-verbatim translation.
In our  implementation, 
negation and the typical infix operators \texttt{AND} and \texttt{OR}
get printed out with parenthesis, to ensure the right precedence -- the
Prolog side ignores superfluous parens anyway.

\subsubsection{Parsing Prolog output}\label{parsing-prolog-output}

Custom parsing of Prolog output provides a nice special case: At the 
beginning of an output block, one or more lines containing a string beginning with 
the sequence \texttt{<html>} are recognized and displayed verbatim; this way
a human-readable version of the derivations can be created in Prolog by nesting \texttt{<div>}s as necessary.
Some CSS provided by the POC is then used by the browser to provide a nice visual display.
Regular prolog output is parsed via the \texttt{ESRAP} library.

\section{Conclusion and Outlook}
\label{conclusion}

We presented \emph{Grants4Companies}, an application in the Austrian
public administration, which uses declarative methods to recommend
business grants based on the data available for the businesses from
sources in the public administration. We also presented the Proof of
Concept implementation of logical reasoning over the formalised grant
conditions, implemented in Common Lisp and Scryer Prolog. A main
interest here lies in the fact that the PoC implementation uses
declarative and logical methods in the context of an application,
which is already live in public administration.

In terms of future work we are steadily extending the list of covered
grants, and are considering automatised rules extraction methods
(e.g.,~\cite{Kovacs:2022Potato,Recski:2021ASAIL}) for speeding up this
process. Extending the coverage of the grants will necessitate the
extension of the logical language and hence the reasoning mechanisms
to further concepts and also towards (limited) reasoning with natural
numbers. We envisage the resulting tool to become a possible basis for
systematic analyses of the Austrian landscape of business grants by
stakeholders in funding agencies and public administration.

%
%
%
 \bibliographystyle{splncs04}


\begin{thebibliography}{10}
\providecommand{\url}[1]{\texttt{#1}}
\providecommand{\urlprefix}{URL }
\providecommand{\doi}[1]{https://doi.org/#1}

\bibitem{Beckert:1996}
Beckert, B., Posegga, J.: Logic programming as a basis for lean automated
  deduction. The Journal of Logic Programming  \textbf{28}(3),  231--236
  (1996). \doi{https://doi.org/10.1016/0743-1066(96)00054-4},
  \url{https://www.sciencedirect.com/science/article/pii/0743106696000544}

\bibitem{belzer1978encyclopedia}
Belzer, J., Holzman, A., Kent, A.: Encyclopedia of Computer Science and
  Technology: Volume 10 - Linear and Matrix Algebra to Microorganisms:
  Computer-Assisted Identification. Taylor \& Francis (1978)

\bibitem{Kleene:1952fk}
Kleene, S.C.: Introduction to Metamathematics. North-Holland, Amsterdam (1952)

\bibitem{Knuth:1984CJ}
Knuth, D.: Literate programming. The Computer Journal  \textbf{27}(2),  97--111
  (1984)

\bibitem{Kovacs:2022Potato}
Kov\'{a}cs, A., G\'{e}mes, K., Ikl\'{o}di, E., Recski, G.: Potato: Explainable
  information extraction framework. In: Proceedings of the 31st ACM
  International Conference on Information \& Knowledge Management. p.
  4897–4901. CIKM '22, Association for Computing Machinery, New York, NY, USA
  (2022). \doi{10.1145/3511808.3557196},
  \url{https://doi.org/10.1145/3511808.3557196}

\bibitem{Mowbray:2023CLSR}
Mowbray, A., Chung, P., Greenleaf, G.: Representing legislative rules as code:
  Reducing the problems of 'scaling up'. Computer Law \& Security Review
  \textbf{48},  105772 (2023).
  \doi{https://doi.org/10.1016/j.clsr.2022.105772},
  \url{https://www.sciencedirect.com/science/article/pii/S0267364922001157}

\bibitem{multlog:stongKleene}
Multlog: Analytic proof systems for strong {K}leene logic ${K}_3$ [pdf
  generated by {MULTLOG}, v.1.16a, \url{https://logic.at/multlog} (2022),
  \url{https://logic.at/multlog/kleene.pdf}

\bibitem{Recski:2021ASAIL}
Recski, G., Lellmann, B., Kov{\'a}cs, A., Hanbury, A.: Explainable rule
  extraction via semantic graphs. In: ASAIL~2021. pp. 24--35. CEUR Workshop
  Proceedings (2021)

\bibitem{Troelstra:2000fj}
Troelstra, A.S., Schwichtenberg, H.: Basic Proof Theory, Cambridge Tracts In
  Theoretical Computer Science, vol.~43. Cambridge University Press, 2 edn.
  (2000)

\bibitem{GDPR}
Regulation (EU) 2016/679 of the European Parliament and of the
  Council of 27 April 2016 on the protection of natural persons with regard to
  the processing of personal data and on the free movement of such data, and
  repealing Directive 95/46/EC (General Data Protection Regulation). OJL
  (2016)

\end{thebibliography}
%
\end{document}